\documentclass[twocolumn,english,prb,showpacs]{revtex4}
\usepackage{epsfig}
\usepackage{graphicx}
\usepackage{amsfonts}
\usepackage[figuresright]{rotating}
\usepackage{amssymb}
\usepackage{amsmath}
\usepackage{amsmath}
\usepackage{graphicx}
\usepackage{hyperref}
\usepackage{color}
\usepackage{soul}

\begin{document}

\title{Correlation effects in pyrochlore iridate thin films grown along the [111] direction}
\author{Qi Chen}
\email[]{chenqi0805@gmail.com}
\affiliation{Department of Physics, The University of Texas at
Austin, Austin, TX, 78712, USA }
\author{Hsiang-Hsuan Hung}
\affiliation{Department of Physics, The University of Texas at
Austin, Austin, TX, 78712, USA }
\author{Xiang Hu}
\affiliation{Department of Physics, The University of Texas at
Austin, Austin, TX, 78712, USA }
\author{Gregory A. Fiete}
\affiliation{Department of Physics, The University of Texas at
Austin, Austin, TX, 78712, USA }

\begin{abstract}
Over the past few years bulk pyrochlore iridates of the form  $A_2$Ir$_2$O$_7$ (where $A$ is a rare earth element, Ir is iridium, and O is oxygen) have been studied as model systems for investigating the interplay of electronic correlations and strong spin-orbit coupling, particularly with the aim of finding correlation-driven topological phases.  In this work,  we use cellular dynamical mean field theory (CDMFT) to study effects of electronic correlations beyond Hartree-Fock theory in thin films of pyrochlore irradiates grown along the $[111]$ direction.  We focus on the bilayer and trilayer systems, and compute the phase diagrams of these systems as a function of electron-electron interaction strength, which is modeled by an on-site Hubbard interaction. By evaluating the $Z_2$ invariant and Chern number using formulas based on the single-particle Green's function and the quasi-particle effective Hamiltonian, we show that on-site correlations can drive an interaction-induced topological phase transition, turning a time-reversal invariant topological insulator and a nearly flat band metal to a correlated Chern insulator (CI) in bilayer and trilayer systems, respectively. By comparing with the Hartree-Fock results, the CDMFT results show that quantum fluctuations enhance the robustness of the interaction-driven CI phase in the thin films. Furthermore, our numerical analysis of the quasiparticle spectrum reveals  that the topological phases we find in our many-body calculations are adiabatically connected to those in the single-particle  picture.
\end{abstract}


\maketitle

\section{Introduction}

Over the past decade, time-reversal invariant topological insulators (TI) in two and three dimensions have received significant theoretical and experimental attention.\cite{TI2010rmp,Z2prl2005,QSHE2005,3DTI2007,Z2Moore2007,Royprb2009,QXLrmp2011,Konig2008,KonigScience2007}
Recently, significant effort has been made to investigate the role of electron-electron interactions in topological states of of matter.\cite{Rachel2010,RanYing2011a,Fa2011, Zhengdong2011,Hohenadler2012,Hung2013,Hung2014,William2014} Though it is generally understood that topological phases described by electronic band structure are robust to weak electron-electron interactions, the full many-body problem of a strongly interacting system remains far from completely understood.\cite{Maciejko:np14,William2014}  One possible effect of electronic interactions is magnetic order that carries with it a topological phase transition (either from a non-topological system to a topological one, or from a TI to a non-topological magnetically ordered state).\cite{He2011,Yoshida2012,He2012,Yoshida2013,Hohenadler2013}

Among the real-material proposals for systems expected to exhibit topological properties, transition metal oxides (TMO) are a promising candidate for realizing nontrivial {\em interacting topological phases}.\cite{Shitade2009,Pesin2010,YBKim2010,Mehdi2011,Ara2012,RanYing2011b,William2012,William2013,Okamoto2014,Nagaosa2014,Kargarian:prl13,Maciejko:prl14,Ruegg:prl12}  Among the bulk (as apposed to thin-film) TMO that have been theoretically studied in the context of topological phases, the irradiates have been particularly singled out.\cite{Shitade2009,Pesin2010,YBKim2010,Mehdi2011,Ara2012,RanYing2011b,William2012,William2013,William2014,Okamoto2014,Nagaosa2014,Kargarian:prl13,Maciejko:prl14,Wan2011}  In these materials, the iridium 5$d$ electrons typically dominate the states near the Fermi energy.  The $d$-orbitals are usually correlated and the large atomic number of iridium means that spin-orbit coupling can be significant as well.  Thus, the iridates are a natural platform for investigating the interplay of strong spin-orbit coupling and correlation effects.\cite{William2014}

\begin{figure}[!htb]
\centering
\epsfig{file=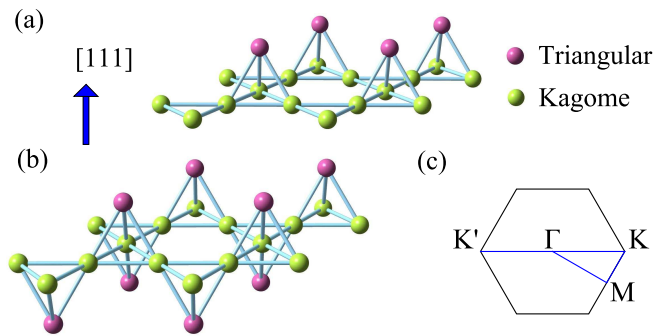,width=1.0\linewidth}
\caption{(Color online) The lattice structure of the (a) bilayer (TK) and (b) trilayer (TKT) pyrochlore iridate thin films grown along the [111] direction. Each lattice site denotes an Ir$^{4+}$ atom. The light green dots form a coplanar kagome lattice, whereas the pink dots form a coplanar triangular lattice. (c) shows the hexagonal Brillouin zone of the TK and TKT systems.}
\label{fig:lattice}
\end{figure}

In addition to the search for three-dimensional topological phases in bulk TMO oxides, thin films (as thin as a bilayer) of TMO have also received significant attention recently.\cite{Fiete:jap15} A number of these studies have focused on the perovskite structure ABO$_3$ (where A is a rare earth element, B is a transition metal, and O is oxygen) where thin films are grown along the [111] direction.\cite{Andreas2011,Andreas2012,Andreas_dist:prb13,RanYing2011a,DiXiao2011,RanYing2011b,Fa2011,Liang:njp13,Okamoto2014,Lado:prb13,Doennig:prb14,Wang:prb15}
In this work, we are interested in thin films of the pyrochlore irididate $A_2$Ir$_2$O$_7$ (where A is a rare earth element, Ir is iridium, and O is oxygen) grown along the [111] direction.\cite{Xiang2012,XiangDFT,Yang:prl14}  The thin film geometry of $A_2$Ir$_2$O$_7$ allows one to investigate the effects of reduced dimensionality on the interplay of correlations and strong spin-orbit coupling.  Along the [111] direction (or equivalent directions), Ir$^{4+}$ ions form alternating layers of kagome and triangular lattices.  See Fig.\ref{fig:lattice} for a visualization of the geometry of the Ir atoms in the bilayer and trilayer geometry.


Our main objective in this work is to go beyond the Hartree-Fock (HF) approximation\cite{Xiang2012,XiangDFT} in the study of the effects of reduced dimensionality on the interplay of correlations and strong spin-orbit coupling in [111] grown pyrochlore iridate thin films.  The intrinsic spin-orbit coupling of the Ir local moments generally breaks the spin-rotational invariance\cite{Pesin2010,William2012} and therefore opens the possibility of long-range magnetic order at non-zero temperatures.  Therefore, we do not expect {\em qualitative} changes to the magnetic order upon the inclusion of spatial fluctuations on top of the magnetic order predicted in HF theory.  On the other hand, we wish to better understand how the quantum fluctuations in time may impact the HF predictions of interacting topological phases (namely the Chern insulator) in these systems.  In particular, we would like to know whether the temporal fluctuations can be so severe that interaction-generated topological phases predicted by HF theory disappear.  For Chern insulators induced ``purely" by electron-electron interactions (that is, the underlying Hamiltonian has {\em no} intrinsic spin-orbit coupling at the non-interacting level), it appears that temporal fluctuations can indeed destroy the topological phase.\cite{Daghofer:prb14}  In our model--that does include a finite spin-orbit coupling at the non-interacting level--we reach the opposite conclusion:  Temporal fluctuations do not generally destroy an interaction-driven topological phase in two-dimensions.

To investigate the physics of temporal fluctuations in our thin-film systems, we employ cellular dynamic mean-field theory (CDMFT).\cite{CDMFT2001,QCTrmp2005}  We focus on two thin film systems of $A_2$Ir$_2$O$_7$ grown along the [111] direction: a bilayer consisting of a kagome and a triangular lattice (KT), and a trilayer consisting of triangular-kagome-triangular (TKT) lattices, as shown in Fig.~\ref{fig:lattice}(a) and (b).

Because our interest in this work is primarily the investigation of temporal fluctuations on interaction-driven topological phases in two dimensional models (as opposed to a first-principles prediction of interaction effects in real material systems\cite{XiangDFT}), we will simplify the problem by restricting the Hilbert space of our Hamiltonian to the ``$J_{\mathrm{eff}}=1/2$" model, appropriate for the limit of a large spin-orbit coupling.\cite{William2014}  While the large spin-orbit coupling limit may not be reached in real materials,\cite{XiangDFT,Wan2011} we expect the central physical results of our study to be unchanged upon the inclusion of the full $t_{2g}$ sub-space of the iridium 5$d$ orbitals.  The reduced Hilbert space allows us to carry out the CDMFT study; the inclusion of the full $t_{2g}$ sub-space is beyond the numerical capabilities presently available.
Because the bandwidth of the states near the Fermi energy is rather narrow, even a moderate interaction can put the system in a strongly correlated regime.

Our CDMFT results show that the correlation effects can drive a topological phase transition from a TI to a magnetic Chern insulator (CI) with nontrivial Chern number $C=1$ in the bilayer (KT) system. In the trilayer (TKT), we find that many-body interaction effects on the nearly flat band near the Fermi surface can induce a topological phase
transition from a nonmagnetic conductor (C) to an interacting CI with Chern number $\pm 1$. Both thin film cases show that the ground state is a trivial magnetic insulator in the strong interaction limit. Compared to the previous HF results,\cite{Xiang2012} our results suggest that moderate quantum fluctuations (captured by CDMFT) can stabilize the interaction-driven CI phases in both bilayer and trilayer systems, effectively enlarging the parameter space where they appear.

Our paper is organized as follows. In Sec.\ref{sec:Hamiltonian}, we first introduce a simplified model Hamiltonian for the thin films of pyrochlore iridates $A_2$Ir$_2$O$_7$. In Sec.\ref{sect:cdmft} we briefly describe the CDMFT method, and explain how to evaluate the Chern number and the $Z_2$ invariant. In Sec. \ref{sect:numericalresult}, we present our numerical results for the bilayer and trilayer cases with the exhaustive phase diagrams and corresponding magnetic configurations.  Finally, we summarize our work in Sec. \ref{sect:summary}.


\section{Model Hamiltonian}
\label{sec:Hamiltonian}
In the pyrochlore oxides $A_2B_2$O$_7$, with $A$ a rare earth element such as Y or La, the $A$-site has a vanishing magnetic moment and the physics is dominated by the transition metal $B$ ions. In this work, we apply the model Hamiltonian proposed for the three-dimensional materials\cite{Pesin2010,William2012} to the quasi-two-dimensional thin film systems.\cite{Xiang2012}  Effectively integrating out the oxygen orbitals involved in indirect hopping processes between $B$-site ions, one obtains a model that only involves the $B$-sites. We will use such a model in this work. The $5d$ atomic orbitals in Ir$^{4+}$ are subject to a cubic crystal field, which splits the $5d$ orbitals into $e_g$ and $t_{2g}$ manifolds. The $e_{g}$ manifold typically lies 2-3 eV above the $t_{2g}$ manifold.\cite{PhysicsofTMO}  Spin-orbit coupling (SOC)  further splits the $t_{2g}$ manifold  into a $J_{\mathrm{eff}}=1/2$ doublet and a $J_{\mathrm{eff}}=3/2$ quadruplet. At infinite SOC strength, the $J_{\mathrm{eff}}=1/2$ states form a low energy manifold (the $J_{\mathrm{eff}}=3/2$ levels are inactive since they are  fully occupied below the Fermi level) for the thin films.

In the pyrochlore iridates $A_2$Ir$_2$O$_7$,  four Ir$^{4+}$ ions form a tetrahedron in the unit cell of a face-centered cubic Bravais lattice and the $J_{\mathrm{eff}}=1/2$ manifold is half-filled.\cite{William2014} In addition, we include an on-site Coulomb repulsion within the $J_{\mathrm{eff}}=1/2$ pseudospin space to obtain the effective Hamiltonian for our thin films:\cite{William2012,Ara2012}
\begin{eqnarray}
  H & = & \sum_{\langle \mathbf{R}_i,\mathbf{R}_{j} \rangle,\sigma \sigma'} ([T_{\mathrm{oxy}}]^{i j}_{\sigma \sigma'}+[T_{\mathrm{dir}}]^{i j}_{\sigma \sigma'})c^{\dagger}_{\mathbf{R}_i \sigma} c_{\mathbf{R}_j \sigma'} \\ \nonumber
   & & -\mu \sum_{\mathbf{R}_i,\sigma} c^{\dagger}_{\mathbf{R}_i \sigma} c_{\mathbf{R} i \sigma} + U \sum_{\mathbf{R}_i} n_{\mathbf{R}_i \uparrow} n_{\mathbf{R}_i \downarrow}  \\ \nonumber
\label{eqn:Ham}
\end{eqnarray}
where $c_{\mathbf{R}_i \sigma}$ annihilates an electron with pseudospin $\sigma$ at the $i$th site of the Bravais lattice vector $\mathbf{R}$. The site index $i$ runs from 1 to 4 in the bilayer KT lattice [in Fig. \ref{fig:lattice}(a)], while it runs from 1 to 5 in the trilayer TKT lattice within a single unit cell [in Fig. \ref{fig:lattice}(b)]. The hopping parameter $T_{\mathrm{oxy}}$ arises from the oxygen-mediated hopping between nearest neighbor (NN) Ir$^{4+}$ atoms with amplitude $t \propto V_{\mathrm{pd}}^2/ \Delta$, where $V_{\mathrm{pd}}$ is the tunneling amplitude between $p$-orbitals of the oxygen and $d$ orbitals from Ir; $\Delta$ is the energy difference between the two atomic orbital levels. The hopping parameter $T_{\mathrm{dir}}$ is the direct NN Ir-Ir hopping due to the direct overlap between the extended $5d$ orbitals, which depends on $t_{\sigma}$ and $t_{\pi}$ from $\sigma$-$\pi$ bonding between the $d$-orbitals.\cite{William2012} The chemical potential $\mu$ is adjusted so that at half filling, each site is has one particle on average in the $J_{\mathrm{eff}}=1/2$ bands. The values of the tight binding parameters in the matrices $[T_{\textrm{oxy}}]$ and $[T_{\textrm{dir}}]$ have been explicitly expressed in Appendix A of Ref.~[\onlinecite{Xiang2012}]. For simplicity,  we choose hopping between oxygen $p$ orbitals and iridium $d$ orbitals, $t_{\mathrm{pd}}=1$, as the unit of energy, and set $t_{\pi}=-2t_{\sigma}/3$ for both bilayer and trilayer system in order to compare with bulk systems\cite{William2012,Ara2012} and thin film results.\cite{Xiang2012}

Following earlier work on pyrochlore iridiates,\cite{William2012,Xiang2012} we select different values of the tight-binding hopping parameter $t_{\sigma}$ to explore different regimes of the parameter space potentially relevant to experimental systems. Because of the underlying triangular lattice in the (111) planes, the Brillouin zones of both the KT and the TKT systems are hexagonal--see Fig. \ref{fig:lattice}(c).

\begin{figure}[!t]
\epsfig{file=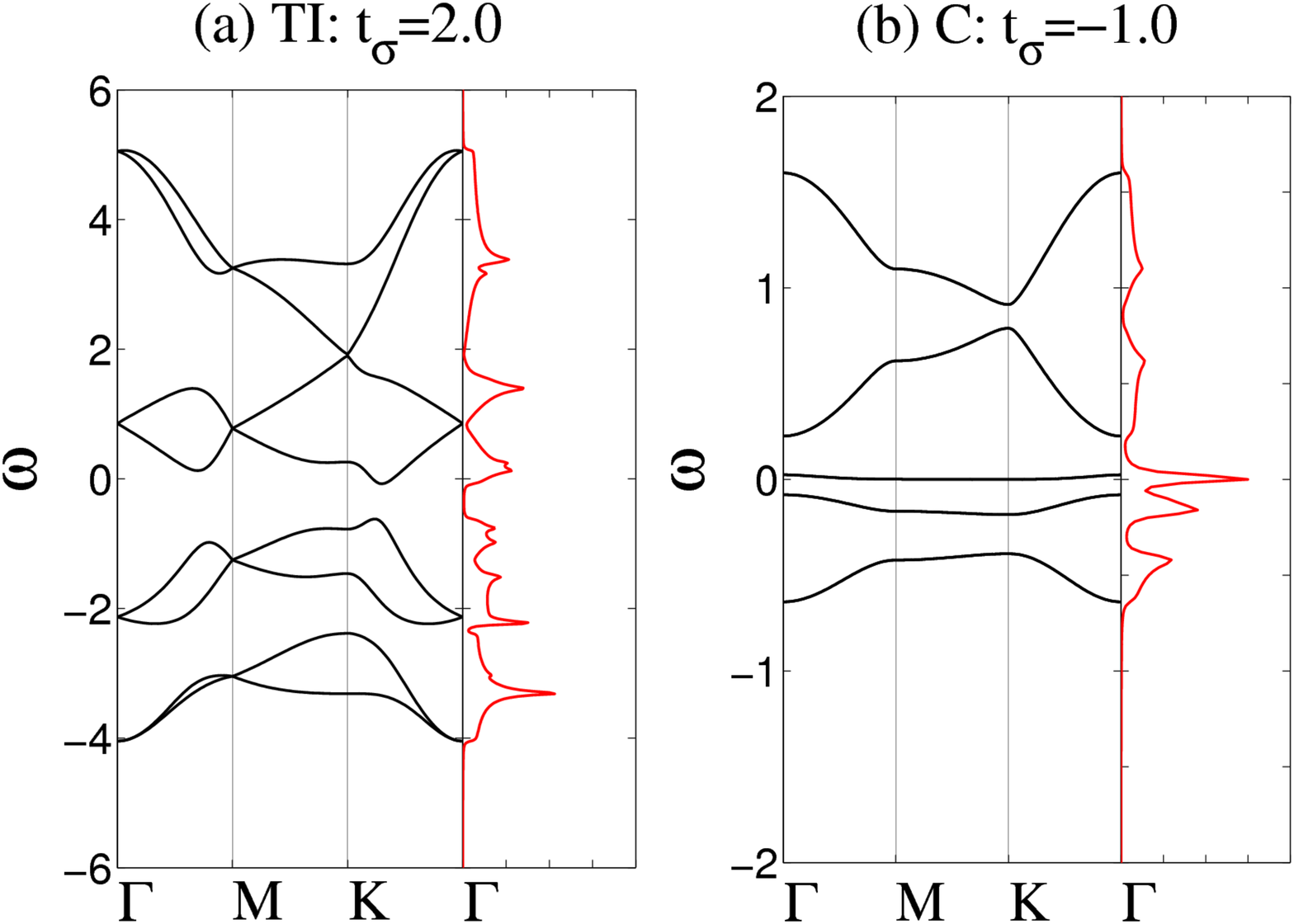,clip=0.1,width=0.9\linewidth,angle=0}
\caption{(Color online) The noninteracting band structure along high symmetry directions in the Brillouin zone and density of states (right side of each figure shown in red) for (a) bilayer (KT) at $t_{\sigma}=2$;
and (b) trilayer (TKT) at $t_{\sigma}=-1$. At half filling, (a) is insulating whereas (b) is metallic.}
\label{fig:spectralKTU0}
\end{figure}

For the bilayer system, we choose $t_{\sigma}=2$ and present the noninteracting band structure and density of states (DOS) in Fig. \ref{fig:spectralKTU0}(a). The KT lattice breaks inversion symmetry and the noninteracting band structure has an energy gap around the Fermi level at half filling. A Kramers degeneracy is found at time reversal invariant momenta (TRIM), i.e., $\Gamma$ and M in Fig. \ref{fig:lattice}(c). By calculating the $Z_2$ invariant using the Fu-Kane formula\cite{3DTI2007} or with Fukui's method,\cite{FukuiZ2} one finds the system is a topological band insulator for $U=0$.\cite{Xiang2012}  For this value of $t_{\sigma}$, we can thus calculate  within the CDMFT formalism how interactions drive the system away from a $Z_2$ TI as the parameter $U$ is increased.

On the other hand, we choose $t_{\sigma}=-1$ for the trilayer case. In this parameter regime, the non-interacting band structure, depicted in Fig. \ref{fig:spectralKTU0}(b), shows nearly degenerate flat bands close to the Fermi level, such that at half filling it is metallic. As is revealed by previous HF results,\cite{Xiang2012} however, a small but finite interaction $U$ will open a gap, and drive the system to a CI.  We wish to study the fate of this transition within the CDMFT formalism.


\section{CDMFT formalism and evaluation of topological invariants}
\label{sect:cdmft}

CDMFT is an extension of the single-site dynamical mean-field theory method\cite{DMFT1996} to include spatial correlation effects within a super-lattice unit cell.\cite{CDMFT2001,QCTrmp2005} As a result, this method can capture the short-distance effect of quantum fluctuations neglected in single-particle approximation and has already been successfully applied to study other complex oxides\cite{Okamoto2014, Andreas2014, Lombardo1996, Phillips2003} and the three-dimensional bulk (and slab) pyrochlore iridates.\cite{Ara2012}  In general, the dynamical mean-field theory method reduces lattice problems with infinite degree of freedom to a type of Anderson impurity problem, in which a cluster of $N_c$ sites hybridizes with electron bath sites (environment):
\begin{eqnarray}
\label{eq:H_imp}
H_{imp}&=&\sum_{\mu\nu\sigma}E_{\mu\nu}^{\sigma \sigma'}c^{\dagger}_{\mu \sigma} c_{\nu \sigma'} + U \sum_{\mu} n_{\mu \uparrow} n_{\mu \downarrow} \nonumber \\
&&+\sum_{\mu l \sigma \sigma'}(V_{\mu l}^{\sigma \sigma'}a^{\dagger}_{l \sigma'}c_{\mu \sigma}+V_{\mu l}^{\sigma \sigma' *}c^{\dagger}_{\mu \sigma}a_{l \sigma'}) \nonumber \\
&&+\sum_{l \sigma} \epsilon_{l\sigma} a_{l\sigma}^{\dagger} a_{l\sigma},
\end{eqnarray}
where the greek symbols $\mu, \nu=1,N_c$ label cluster sites; $l=1,\ldots,N_b$ labels bath sites, and $\sigma, \sigma'$ are pseudo-spin labels. The electron operators $c^{\dag}_{\mu \sigma}$ ($c_{\mu \sigma}$) apply to the cluster sites, whereas $a^{\dag}_{l\sigma}$ ($a_{l\sigma}$) to the bath sites. The hopping integrals and chemical potential
within the cluster are incorporated by the matrix elements of $E$, which are obtained from the tight-binding parameters $[T_{\textrm{oxy}}]$ and $[T_{\textrm{dir}}]$. The $V$'s and $\epsilon$'s are bath parameters describing hybridization between the clusters and bath sites, and on-site energy levels, respectively.  The values of the $V$, the $\epsilon$ and the lattice Green's functions of Eq.(\ref{eq:H_imp}) are numerically determined via an exact diagonalization and self-consistency procedure.

The self-consistency procedure is as follows: an initial input of the bath parameters $V$ and $\epsilon$
is given in Eq.(\ref{eq:H_imp}) to solve the impurity Hamiltonian. From the cluster impurity Hamiltonian, we compute the cluster
Green's function $\hat{G}(i\omega)$ as a $2N_c \times 2N_c$ (number of sites $=N_c$ and number of pseudospin degrees of freedom $=2$) matrix, as well as the cluster self-energy
\begin{eqnarray}
\hat{\Sigma}_c(i\omega)=[\hat{\mathcal{G}}(i\omega)]^{-1}-[\hat{G}(i\omega)]^{-1},
\end{eqnarray}
 where $\hat{\mathcal{G}}$ is the non-interacting cluster Green's function. In CDMFT, it serves
as the Weiss field describing the coupling of the cluster to the environment.
Then the lattice Green's function is coarse-grained as
\begin{eqnarray}
\label{eq:G_loc}
\hat{G}_{\mathrm{loc}}(i \omega)=\frac{N_c}{N}\sum_{\mathbf{k} \in \mathrm{BZ}}\Big[(i \omega+\mu)\mathbf{\hat{1}}-\hat{t}(
\mathbf{k})-\hat{\Sigma}_c(i\omega)\Big]^{-1},
\end{eqnarray}
where $\mathbf{k}$ is the momentum in the Brillouin zone, and $\hat{t}(\mathbf{k})$ is the Fourier transformed hopping integral.
The self-consistent loop is closed by calculating the new Weiss function
\begin{eqnarray}
\label{eq:Weiss_new}
[\hat{\mathcal{G}}_{\mathrm{new}}(i\omega)]^{-1}&=&[\hat{G}_{\mathrm{loc}}(i\omega)]^{-1}+\hat{\Sigma}_c(i\omega),
\end{eqnarray}
and requiring $[\hat{\mathcal{G}}_{\textrm{new}}(i\omega)] \simeq [\hat{\mathcal{G}}(i\omega)]$ to within a prescribed accuracy. In updating the $V$ and the $\epsilon$, we fit the new Weiss function on the Matsubara axis with conjugate gradient methods and chose an artificial low temperature $k_B T=0.01$ (in units of $t$) to simulate the zero temperature case.  This value is much smaller than other characteristic energy scales in the problem, including the energy spacing due to the discretization of the momentum space in the Brillouin zone, and thus approximates the zero temperature limit well.

In our work, we chose a tetrahedron as a unit cell cluster for the bilayer case, i.e. $N_c=4$;
while for the trilayer case, the unit cell cluster is chosen as a tetrahedron with an external (pink) site linked to one of its 4 corners, i.e. $N_c=5$ [cf. Fig.~\ref{fig:lattice}(a) and (b)].
In order to solve Eq.(\ref{eq:H_imp}), exact diagonalization with the Lanczos algorithm\cite{DMFT1996} is employed to get the ground state properties of the cluster, such as the
magnetic order parameter and single-particle Green's function. In the bilayer case, we compare $N_b=4$ and $N_b=8$ to determine how the phase diagram depends on the finite bath sites\cite{Ishida2010}; In the trilayer case, we choose $N_b=5$.

Next, we briefly describe how we numerically calculate topological invariants using the CDMFT method. Interested readers can see more details in Appendix \ref{sec:topological hamiltonian}.  In a non-interacting topological band insulator, the Chern number is equal to the TKNN number\cite{TKNN}
\begin{eqnarray}
C &=& \int \frac{d^2k}{2\pi} \sum_{\alpha=\mathrm{filled}} f_{xy},
\label{eq:TKNN}
\end{eqnarray}
where $f_{ij}=\partial a_i(\mathbf{k})/\partial k_j-\partial a_j(\mathbf{k})/\partial k_i$ is the Berry curvature, and
$a_i(\mathbf{k})=-i\langle u_{\alpha}(\mathbf{k}) |\frac{\partial}{\partial k_i}| u_{\alpha}(\mathbf{k}) \rangle$ is the Berry connection.
The summation is over filled band indicies.  Here, $|u_{\alpha}(\mathbf{k})\rangle$ denotes the $\alpha$-th eigenstate of the noninteracting Hamiltonian $H(k)$.
There is no inversion symmetry in the pyrochlore oxide [111] grown bilayer system. To calculate the $Z_2$ index, $(-1)^{P_2}$, in a generic way, we resort to the approach given by Fukui\cite{FukuiZ2} in the half-BZ,
\begin{eqnarray}
\label{eq:obstruction to Stokes_1}
P_2 &=& \int_{\mathrm{BZ}} d^2 k f_{12}-\int_{\partial \mathrm{BZ}} d \mathbf{k} \cdot \mathbf{a}\ \mathrm{mod}\ 2.
\end{eqnarray}
Note that in the $Z_2$ formalism, a time-reversal smooth gauge has to be chosen as
\begin{equation}
\label{eq:TR smooth}
| u_{\alpha}(-\mathbf{k}) \rangle=T | u_{\alpha}(\mathbf{k}) \rangle,
\end{equation}
where $T$ is the time-reversal operator.

For a general interacting system, one is no longer able to use the non-interacting invariants given in Eq.\eqref{eq:TKNN} and Eq.\eqref{eq:obstruction to Stokes_1}. Instead, one must express the invariant in terms of the single-particle Green's function at zero frequency.\cite{Zhong2010,Zhong2013} With the CDMFT method, we can easily obtain the zero-frequency Green's function $\hat{G}(i\omega=0,\mathbf{k})$ at self-consistency. The zero-frequency Green's function allows one to define a ``topological Hamiltonian"\cite{Zhong2010,Zhong2013} from which one can compute the corresponding invariant,
\begin{equation}
\label{eq:ham_top}
H_{\mathrm{top}}(\mathbf{k})=-\hat{G}^{-1}(0,\mathbf{k}).
\end{equation}
One diagonalizes the ``topological Hamiltonian" to obtain the eigenvalues
\begin{equation}
\label{eq:ham_top_eig}
H_{\mathrm{top}} |\alpha,\mathbf{k} \rangle = \mu_{\alpha}(\mathbf{k}) |\alpha,\mathbf{k} \rangle,
\end{equation}
where $\alpha$ is a ``band" index. Filled ``bands" are selected as eigenvalues with $\mu_{\alpha}(\mathbf{k})<0$. Then the Berry connections $a_i(\mathbf{k})$ are constructed by replacing
the non-interacting Hamiltonian eigenstates $| u_{\alpha}(\mathbf{k})\rangle$ with the topological Hamiltonian eigenstates $|\alpha, \mathbf{k}\rangle$, and plugging the result into Eq.(\ref{eq:TKNN}) and Eq.(\ref{eq:obstruction to Stokes_1}) to evaluate the Chern number and $Z_2$ invariant, respectively.

For weak to moderate strength interactions, quasiparticles exists with a finite lifetime. Because of the lifetime broadening of the states, it can be difficult to determine if a gap exists by directly evaluating the interacting  quasiparticle spectral function $A(\omega,\mathbf{k})=-\frac{1}{\pi}\textrm{Im}G(\omega+i\eta,\mathbf{k})$: The quasiparticle bands will have a finite width in energy and are sensitive to the value of $\eta$. In order to deal with this numerical issue, we determine the ``effective" quasi-particle dispersion from the quasiparticle effective Hamiltonian Eq.(\ref{eq:quasi-dispersion-CDMFT}) by expanding the local self-energy up to first order in frequency. The details are explained in Appendix \ref{sec:quasiham}. We emphasize that this effective Hamiltonian mainly captures the quasi-particle features around the Fermi energy ($\omega=0$).


\section{Numerical Results}
\label{sect:numericalresult}

In this section, we present our CDMFT results. In the HF theory, the single-particle band structure is renormalized by the Hartree and Fock terms in the mean-field Hamiltonian,\cite{Xiang2012} and the self-energy of the single-particle Green's function depends only on the momentum $k$ (being independent of the frequency $\omega$). Therefore, we re-examine the fate of topological phases and transitions obtained by HF with CDMFT. In the latter approach, quantum fluctuations and correlations are included non-perturbatively in the self-energy within the cluster. However, in CDMFT calculations, one usually obtains a hysteresis behavior when the phase transition is of the first order. Namely, the critical value of $U$ (say $U_{c1}$) when the system undergoes a transition from phase $A$ to phase $B$ by increasing $U$ does not coincide with the one (say $U_{c2}$) when the system undergoes a transition from phase $B$ to phase $A$ by decreasing $U$. In order to determine the complete phase diagram, we numerically calculated and compared the free energy per site when both magnetic and non-magnetic solutions coexist for a particular value of $U$.  The phase is assigned to the state with the lower free energy.

\subsection{Bilayer system}

\begin{figure}[!t]
\epsfig{file=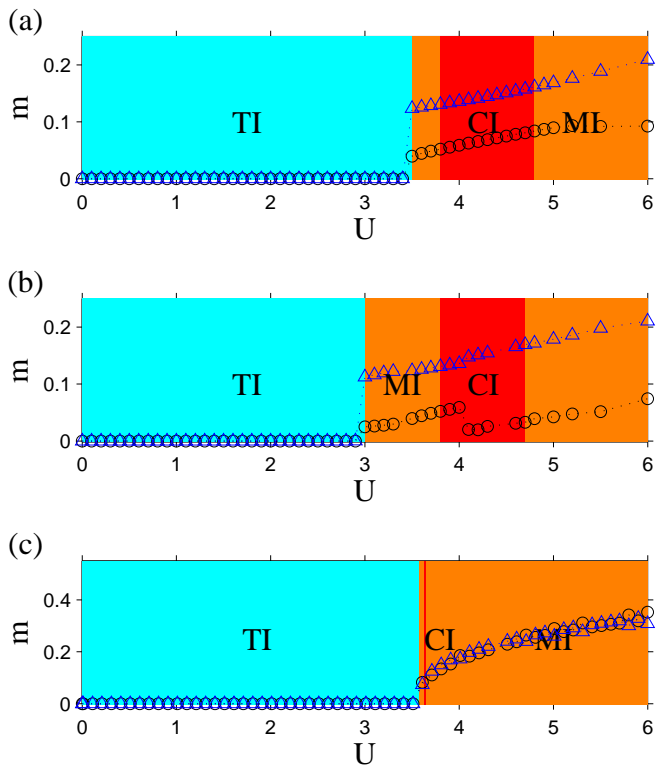,clip=0.1,width=\linewidth,angle=0}
\caption{(Color online) Bilayer (KT) phase diagram for $t_{\sigma}=2$. (a) $N_b=4$ CDMFT results; (b) $N_b=8$ CDMFT results; (c) HF results. In (a), the net magnetic moment defined in the unit cell $m$ (black circles) and averaged magnetic moment per site $\bar{m}$ (blue triangles) are plotted as a function of $U$. TI: time-reversal invariant topological insulator; MI: trivial magnetic insulator; CI: topological Chern insulator.}
\label{fig:phasediagramKTts2}
\end{figure}

The $N_b=4$ and $N_b=8$ CDMFT phase diagram together with the magnetization in the unit cell are presented as a function of $U$ in Fig.~\ref{fig:phasediagramKTts2}(a) and (b), respectively.  While our results are for the parameter $t_{\sigma}=2$, we also confirmed that a similar phase diagram is obtained with $t_{\sigma}=1$. For comparison, we also show the phase diagram obtained by Hartree-Fock theory Fig.~\ref{fig:phasediagramKTts2}(c).

\begin{figure}[!t]
\epsfig{file=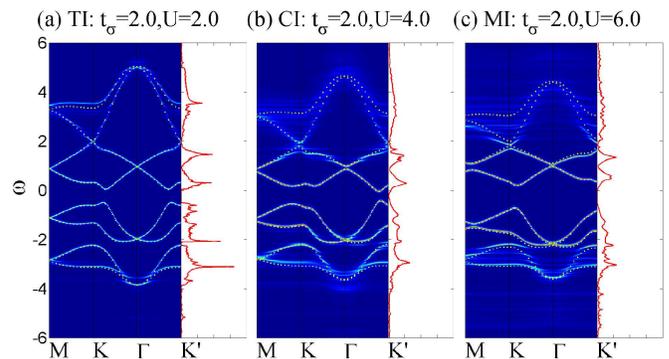,clip=0.1,width=\linewidth,angle=0}
\caption{Spectral weights along high symmetry lines and local density of states for the bilayer (KT) with $t_{\sigma}=2$, $N_b=8$. The broadening factor is $\eta=0.02$. The yellow dots show quasiparticle spectrum obtained from Eq.\eqref{eq:quasi-dispersion-CDMFT}. (a) TI: $U=2.0$; (b) CI: $U=3.0$; (c) MI: $U=6.0$}
\label{fig:spectralKTts2}
\end{figure}

Both the CDMFT and HF approaches show that there exist three different phases upon increasing $U$, and that the phase boundaries are quantitatively similar, except for the region of the Chern insulator (CI) which is much larger when quantum fluctuations are included.
There exists a finite range of values,  $0 \le U\lesssim 3.5$ in $N_b=4$ and $0 \le U\lesssim 3.0$ in $N_b=8$ (CDMFT), for which the system remains in the time-reversal symmetric (TRS) TI phase. The spectral functions and the quasiparticle band structures (see Appendix~\ref{sec:quasiham}) computed from $N_b=8$ CDMFT are plotted in Fig.~\ref{fig:spectralKTts2}(a), while $N_b=4$ gives similar results.
We can see from  Fig.~\ref{fig:spectralKTts2}(a) that the Kramers degeneracy for quasiparticle bands remains intact and the finite band gap leaves the $Z_2$ invariant $\nu=1$ unchanged. In this case, there is no magnetization induced, and the phase is adiabatically  connected to the noninteracting TI phase. Thus, the TI is stable against moderate electronic correlation, as would be expected for a system with a gap in the excitation spectrum about the Fermi energy.

\begin{figure}[!htb]
\epsfig{file=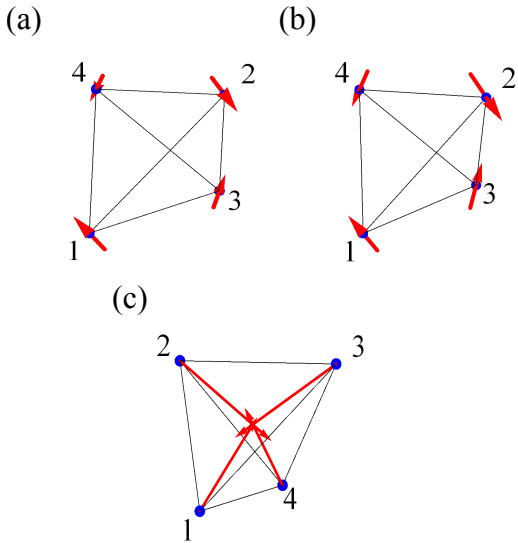,clip=0.1,width=0.8\linewidth,angle=0}
\caption{Magnetic configurations of the bilayer system within a unit cell at $t_{\sigma}=2$ and at (a) $U=4.0$ (CI) and (b) $U=5.0$ (MI). Both magnetic configurations resemble the antiferromagnetic ordering discovered in Ref. [\onlinecite{Ara2012,William2012}]; where moment 1 (2) intersects with moment 4 (3), or  moment 1 (2) intersects with moment 3 (4). To compare with the bulk case,\cite{William2012} (c) takes $t_{\sigma}=-1,U=4.0$ and is a trivial MI. The configuration (c) is close to the ``All-in/All-out" (AIAO) order obtained in the three-dimensional bulk,\cite{William2012} but with a non-vanishing net moment pointing in the (111) direction.  The AIAO order will have zero $m$ but nonzero $\bar{m}$.}
\label{fig:KTmagconfig}
\end{figure}

For $U \gtrsim 3.0$ (in CDMFT), TRS is broken and the system becomes a magnetic insulator. At $U_c \approx 3.0$, a magnetization jump appears, which indicates a first-order phase transition from a non-magnetic to a magnetic insulator. The net magnetic moment, in units of $g \mu_B$ depicted by black circles in Fig.~\ref{fig:phasediagramKTts2}(a), is defined as
\begin{eqnarray}
m=|\sum_i \mathbf{S}_i|,
\end{eqnarray}
where $i$ runs over all sites within a unit cell.
On the other hand, in Fig. \ref{fig:phasediagramKTts2}(a), the averaged magnetic moment per site, defined as
\begin{eqnarray}
\bar{m}=\frac{1}{N_c}\sum_i |\mathbf{S}_i|,
\end{eqnarray}
shows a monotonicaly increasing magnetization upon increasing $U$.  A similar behavior is also captured using the HF theory shown in  Fig.~\ref{fig:phasediagramKTts2}(b). In this regime, the ground state is a trivial magnetic insulator (MI) due to the trivial Chern number $C=0$.

Upon increasing $U$, a finite range that harbors the interaction-induced CI phases is observed at $U \thicksim 3.8-4.8$ for $N_b=4$, $U \thicksim 3.8-4.7$ for $N_b=8$ by CDMFT and at $U \thicksim 3.63-3.65$ by HF theory. In comparison between $N_b=4$ and $N_b=8$, it is found that the magnetic phase boundary is shifted by around $0.5$ in $U$. We ascribe this shift to the effect of finite bath sites. The spectral functions and the quasiparticle dispersion for the CI from $N_b=8$ are plotted in  Fig.~\ref{fig:spectralKTts2}(b). To avoid the broadening of the spectral function by the imaginary part of the self-energy, we identify the gap around Fermi level by examining the band gaps from both topological Hamiltonian Eq.(\ref{eq:ham_top}), and the quasiparticle effective Hamiltonian Eq.(\ref{eq:quasi-dispersion-CDMFT}). In both themes, the band topologies are well-defined and a nontrivial Chern number $C=+1$ is found for half-filled bands. One can observe from Fig.~\ref{fig:spectralKTts2} that the quasi-particle dispersion matches well with the ridges of spectral weights along high symmetry lines, according to our definition of quasi-particle effective Hamiltonian Eq.(\ref{eq:quasi-dispersion-CDMFT}). The Kramers degeneracy at the TRIM points is lifted  in the presence of finite magnetic moments. More interestingly, the CI phase ``survives" for wider range of $U$ values in the CDMFT phase diagram than in the HF phase diagram. This observation suggests that quantum fluctuations stabilize the interaction-induced CI phase.

The magnetic configurations for the CI and MI for $N_b=8$ are illustrated in Fig.\ref{fig:KTmagconfig}(a) and (b), respectively. Similar configurations are also found for $N_b=4$. For the parameters we have studied, the magnetic configurations in the bilayer case are similar to that in the bulk pyrochlore oxides,\cite{William2012,Ara2012} denoted as $\Gamma_5$,\cite{DisselerAIAO} except for the non-vanishing net magnetization in the tetrahedron ($m\ne 0$) due to spatial anisotropy between the in-plane and out-of-plane directions. This pattern is called the antiferromagnetic ordering in the previous studies of the bulk system.\cite{Ara2012,William2012} In the CI phase of Fig.~\ref{fig:phasediagramKTts2}(b) ($N_b=8$), we found a discontinuity in the magnitude of net magnetization around $U \thicksim 4.0$ due to a sudden drop of the ratio between the magnitude of moment 3 and moment 1 in Fig. \ref{fig:KTmagconfig}(a), while the orientation of the moments remains almost the same. This phenomenon does not appear in Fig.~\ref{fig:phasediagramKTts2}(a) ($N_b=4$). Due to the finite number of bath sites used in ED solver, it is not clear whether there is indeed a magnetic phase transition in CI phase. But our results indicate that the CI phase is robust against a small change in magnetic order.

It is interesting to compare the discrepancy between the bilayer case and the three-dimensional bulk. To further examine magnetic configurations between the bulk and the bilayer system, we present the magnetic pattern for $t_{\sigma}=-1$  in Fig.~\ref{fig:KTmagconfig}(c). The magnetic moments are aligned in close similarity to the AIAO configuration, denoted as $\Gamma_3$, found in the three-dimensional bulk materials.\cite{DisselerAIAO} The nonzero net moment in the tetrahedron is pointing in the [111] direction due to the failure of cancelation between moment 1 (in triangular layer) and sum of moment 2, 3, and 4 (in the coplanar kagome layer). Since the Chern number $C=0$, it is still a MI.

\begin{figure}[!t]
\epsfig{file=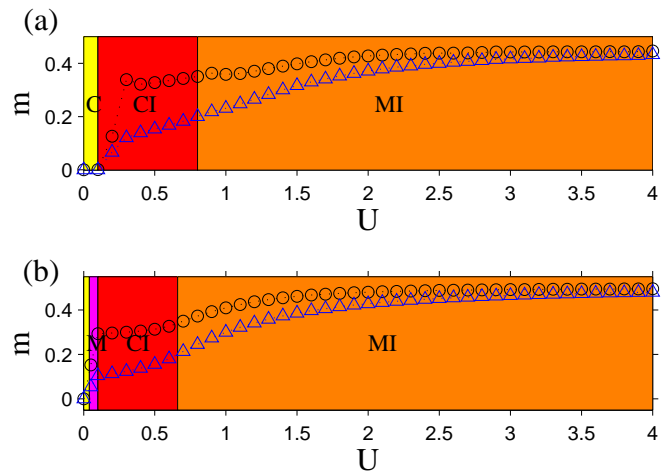,clip=0.1,width=\linewidth,angle=0}
\caption{Trilayer (TKT) phase diagram for $t_{\sigma}=-1$. (a) CDMFT results; (b) HF results. In (a), the net magnetic moment defined in the unit cell $m$ (black circles) and averaged magnetic moment $\bar{m}$ (blue triangles) are plotted as a function of $U$. C: trivial nonmagnetic conductor; M: trivial magnetic conductor.}
\label{fig:phasediagramTKTts-1}
\end{figure}

\subsection{Trilayer system}
The $t_{\sigma}=-1$ noninteracting band structure for the TKT system is shown in Fig. \ref{fig:spectralKTU0}(b). Unlike the bilayer system, the TKT system has both inversion symmetry and TRS. Thus, the noninteracting bands are doubly degenerate and a metallic ground state is expected with a half-filled nearly flat band lying at the Fermi energy. The ground state of the $U=0$ system is a trivial conductor (denoted ``C"). The trilayer phase diagram as a function of the interaction $U$ is shown in Fig.~\ref{fig:phasediagramTKTts-1}.

\begin{figure}[!t]
\epsfig{file=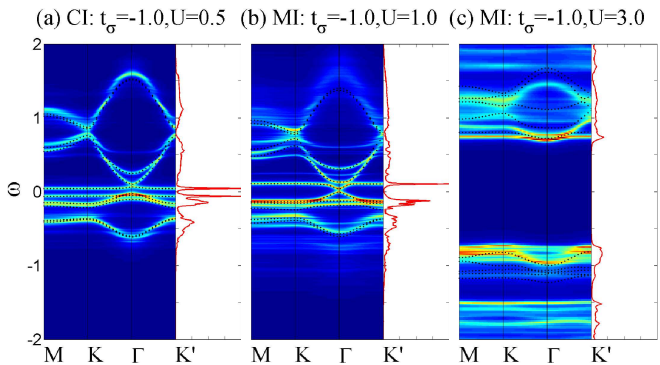,clip=0.1,width=\linewidth,angle=0}
\caption{Spectral weights along high symmetry lines and local density of states for trilayer (TKT) $t_{\sigma}=-1$. The broadening factor is $\eta=0.02$. The black dots show quasiparticle spectrum. (a) CI: $U=0.5$; (b) MI: $U=1.0$; (c) MI: $U=3.0$.}
\label{fig:spectralTKT}
\end{figure}
\begin{figure}[!tb]
\epsfig{file=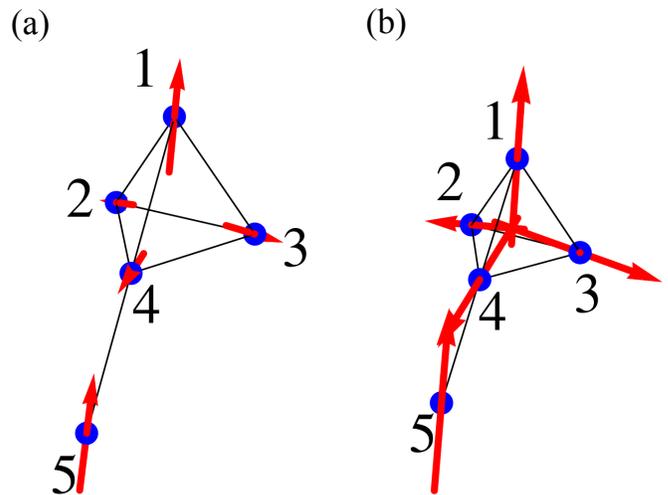,clip=0.1,width=\linewidth,angle=0}
\caption{Magnetic configuration of TKT within a unit cell. (a) $t_{\sigma}=-1,U=0.5$; (b) $t_{\sigma}=-1,U=4.0$; In (a) and (b), the magnetic configuration resembles AIAO but with nonzero net magnetization in a tetrahedron. The magnetic moment 1 and 5 are almost parallel, which enhances the ferromagnetism within a unit cell and the net magnetic flux through the [111] plane.}
\label{fig:TKTmagconfig}
\end{figure}

When the Hubbard interaction reaches $U \approx 0.2$ (in CDMFT), the system breaks TRS  and the doubly degenerate flat bands split into two nearly-flat bands carrying opposite Chern numbers. Therefore, half-filled bands give a nontrivial Chern number $C=\pm1$. Meanwhile, as can be seen in the spectral weight along high symmetry line and local density of states in Fig. \ref{fig:spectralTKT}(a), one can identify a finite gap in this regime. Combining these observation, it is a nontrivial CI. This indicates the single-particle nature of the interaction-induced CI appearing in TKT.
The corresponding magnetic order for TKT is still non-colinear and close to the HF results in both bulk\cite{William2010} and trilayer\cite{Xiang2012} systems, as shown in Fig.~\ref{fig:TKTmagconfig}(a).

If the interaction strength is further increased to $U \approx 0.8$,  a quadratic band touching appears around the $\Gamma$ point and the band topology becomes trivial [cf. Fig.~\ref{fig:spectralTKT}(b)]. Upon further increasing $U$, another topological quantum phase transition occurs. We confirm this by measuring the band gaps around the Fermi level using both the topological Hamiltonian and the quasi-particle Hamiltonian. By diagonalizing the quasi-particle effective Hamiltonian defined in Appendix~\ref{sec:quasiham} and calculating the corresponding Chern number of the filled bands, one finds the same change in the band topology found directly from the corresponding analysis on the topological Hamiltonian determined from the single-particle Green's function.

For $U>0.8$, the TKT system becomes a trivial MI, and the magnetic configuration resembles (but is different from) the AIAO pattern similar to what is found in the bulk. However, for any value of $U$, the CDMFT results do not indicate the presence of the magnetic conductor phase [labeled ``M" in Fig.~\ref{fig:phasediagramTKTts-1} (b)], as predicted by the HF theory. But similarly, the CI phase in the trilayer system is more stable in the CDMFT phase diagram than in the HF phase diagram, showing that moderate correlations and quantum fluctuations tend to favor to the CI phase.

Finally, we note that while in the bilayer system, the CI phase is ``born" out of a parent TI state, the same is not true of the CI in the trilayer: The CI in the TKT system is an example of interactions driving a topological state without a parent topological state.  Combined, the bilayer and trilayer CDMFT results show that the general conclusion reached within the the HF theory--that a CI can emerge from both gapped and gapless states at the non-interacting level--is robust to the inclusion of quantum fluctuations.


\section{Summary}
\label{sect:summary}

In summary, we go beyond the single-electron approximation to re-investigate the phase diagram of bilayer (KT) and trilayer (TKT) thin film pyrochlore iridates in an effective $J_{\mathrm{eff}}=1/2$ model with on-site Hubbard interaction. By applying the cellular dynamical mean-field theory (CDMFT) method with an exact diagonalization cluster impurity solver, the correlation effects in the model Hamiltonian we studied have been fully incorporated in the time domain, and partially in the spatial degrees of freedom.  Importantly, we show that local quantum fluctuations will not destroy the magnetic order and interaction-driven topological phases of these quasi-two-dimensional systems. The effect from different choice of bath levels is summarized in Fig.\ref{fig:phasediagramKTts2} and a comparison of the CDMFT and Hartree-Fock results are shown in Fig.\ref{fig:phasediagramKTts2} and Fig.\ref{fig:phasediagramTKTts-1}.  An interesting result to emerge from this comparison is that in both systems the quantum fluctuations captured in CDMFT (but left out of HF theory) tend to stabilize the Chern insulator phase by enlarging the region of Hubbard $U$ values over which it occupies the phase diagram.  It is worth mentioning that we also studied other values of $t_\sigma$ and in all cases found a similar level of agreement between the CDMFT and HF phase diagrams to the phase diagrams shown in this manuscript.  Thus, we conclude that the HF calculations are reliable for predicting the qualitative features of the phase diagrams of systems similar to those we study here--namely that there is a non-vanishing spin-orbit coupling at the non-interacting level of the Hamiltonian.  In particular, one does not want to rely on a purely interaction-driven spin-orbit coupling to generate the Chern insulator phase.\cite{Daghofer:prb14}

The magnetic order that appears above a critical value of $U$ is another important feature of our results. The configurations of local moments under the CDMFT calculations are non-collinear in both bilayer and trilayer systems and they highly resemble their counterparts in the bulk material with the same tight-binding parameters, though they are different in detail because of the lowered symmetry of the films. For example, the net magnetic moment in the unit cell is non-vanishing due to the quasi-two-dimensional nature of the system, which brings additional anisotropy in the magnetization. Moreover, our CDMFT study indicates that the TI phase exists with an interaction strength sufficient to break the time-reversal symmetry. This correlated topological phase is adiabatically connected to the topological band insulator with the band dispersion and topology given by the ``quasiparticle effective Hamiltonian" and ``topological Hamiltonian", as we define in the text.  It is worth pointing out that although our numerical results suggest positive outcomes for topological phases to be hosted in many-body interacting systems, the energy scales for these phases to be detected in experiments are relatively small--being set by the gap value of the quasi-particle bands. The extent to which the observable topological transport properties survive thermal fluctuations and disorder in interacting phases is an important open question for future work.

\section{Acknowledgment}  We gratefully acknowledge financial support from ARO Grant No. W911NF-14-1-0579, NSF Grant No. DMR-0955778, and
DARPA grant No. D13AP00052. We thank Ara Go for providing us with numerical data and Chungwei Lin, William Witczak-Krempa for helpful discussions. Simulations were performed in the Texas Advanced Computing Center (TACC) in the University of Texas at Austin.  URL:http://www.tacc.utexas.edu  The crystal structure were drawn with Balls and Sticks.\cite{ozawa2004balls}



\appendix

\section{Calculation of Chern number and $Z_2$ invariant}
\label{sec:topological hamiltonian}

In Sec. \ref{sect:cdmft}, we have briefly introduced the approach to calculate the topological invariants using the single-particle Green's function.  Here we provide more details.
In a noninteracting topological band insulator, the Chern number and $Z_2$ invariant are evaluated with the following TKNN\cite{TKNN} and the generic $Z_2$ formula\cite{Z2prl2005}
\begin{eqnarray}
C &=& \int \frac{d^2k}{2\pi} \sum_{\alpha=\mathrm{filled}} f_{xy}, \\
(-1)^{\nu}&=&\prod_{i=1}^{4} \frac{\sqrt{\textrm{Det}[B(\Lambda_i)]}}{\textrm{Pf}[B(\Lambda_i)]},
\label{eq:TKNN and Pfaffian}
\end{eqnarray}
where $f_{ij}= \frac{\partial a_i(\mathbf{k})}{\partial k_j}-\frac{\partial a_j(\mathbf{k})}{\partial k_i}$ and
$a_i(\mathbf{k})=-i\langle u_{\alpha}(\mathbf{k})|\frac{\partial}{\partial k_i}|u_{\alpha}(\mathbf{k}) \rangle$ are the Berry curvature and Berry connection, respectively. Here $|u_{\alpha}(\mathbf{k})\rangle$ are the noninteracting Hamiltonian eigenstates, i.e. $H(\mathbf{k})|u_{\alpha}(\mathbf{k})\rangle=E_{\alpha}(\mathbf{k})|u_{\alpha}(\mathbf{k})\rangle$. The summation is over filled band index for $E_{\alpha}(\mathbf{k})<E_F$.
$B_{mn}(\Lambda_i)=\langle u_{m}(-\mathbf{k})|T|u_{n}(\mathbf{k}) \rangle$
is the sewing matrix entry on the overlap between Bloch state $u_{m}(-\mathbf{k})$ and the Kramer partner of Bloch state $u_{n}(\mathbf{k})$ at TRIM points $\Lambda_i$. In the quasi-two-dimensional thin films, there are four TRIM points: $\Gamma$ and $M_{1,2,3}$.
 An alternative approach to calculate the $Z_2$ index is to see it as an obstruction to Stokes' theorem in the half-BZ,
\begin{eqnarray}
\label{eq:obstruction to Stokes}
P_2 &=& \int_{\mathrm{BZ}} d^2 k f_{12}-\int_{\partial \mathrm{BZ}} d \mathbf{k} \cdot \mathbf{a}\ \mathrm{mod}\ 2,
\end{eqnarray}
where in the Berry connection for $a_i(\mathbf{k})$, a time reversal smooth gauge is obtained by choosing the state and the corresponding time-reversal partner
\begin{equation}
\label{eq:TR smooth}
| u_{n}(-\mathbf{k}) \rangle=T | u_{n}(\mathbf{k}) \rangle,
\end{equation}
where $T$ is the time-reversal operator.
The topological order parameters for interacting topological insulators are derived from topological field theory as
\begin{eqnarray}
\label{eq: C1 and P2}
C_1&=&\frac{i}{24\pi^2} \int d^2 k d\omega \epsilon^{\mu \nu \tau} \nonumber \\
&& \mathrm{Tr} \big[ G \frac{\partial G^{-1}}{\partial q_{\mu}}
G \frac{\partial G^{-1}}{\partial q_{\nu}} G \frac{\partial G^{-1}}{\partial q_{\tau}} \big], \\
P_2 &=& \frac{1}{120} \epsilon^{\mu \nu \rho} \int_{-1}^1 du \int_{-1}^1 dv \int \frac{d^3k}{(2\pi)^3} \nonumber \\
&& \mathrm{Tr} \big[ G \frac{\partial G^{-1}}{\partial q_{\mu}}G \frac{\partial G^{-1}}{\partial q_{\nu}} G \frac{\partial G^{-1}}{\partial q_{\rho}} \nonumber \\
&& G \frac{\partial G^{-1}}{\partial u} G \frac{\partial G^{-1}}{\partial v} \big] \ \mathrm{mod}\ n.
\end{eqnarray}
Note that in the above formulas, one integral is over all frequency range.
It is shown in Refs.[\onlinecite{Zhong2010,Zhong2013}] that the finite-frequency Green's function $G(i\omega,\mathbf{k})$ is topologically equivalent  to the zero-frequency Green's function $G(i\omega=0,\mathbf{k})$. As a consequence,  the above formulas can be simplified by using the zero-frequency Green's function
to reduce the integral by one dimension. Furthermore, one can define a ``topological Hamiltonian",\cite{topoham2013}
\begin{equation}
H_{\mathrm{top}}(\mathbf{k})=-\hat{G}^{-1}(0,\mathbf{k}).
\end{equation}
Since the zero-frequency Green's function is Hermitian, the eigenvalues for the ``topological Hamiltonian" are real
\begin{equation}
\label{eq:ham_top_eig}
H_{\mathrm{top}} |\alpha,\mathbf{k} \rangle = \mu_{\alpha}(\mathbf{k}) |\alpha,\mathbf{k} \rangle,
\end{equation}
where $\alpha$ is a ``band" index. Filled ``bands" have eigenvalues $\mu_{\alpha}(\mathbf{k})<0$. We construct the Berry connection and apply Eq.(\ref{eq: C1 and P2}) and Eq.(\ref{eq:obstruction to Stokes}) with the noninteracting Hamiltonian eigenstates $|u_{\alpha}(\mathbf{k})\rangle$ replaced by topological Hamiltonian eigenstates $|\alpha, \mathbf{k}\rangle$ to evaluate both the first Chern number and $Z_2$ invariant in pyrochlore oxides thin films, which do not have
inversion symmetry in general. The numerical evaluation of the integral over BZ is based on lattice discretization of the gauge field $a_i(\mathbf{k})$.\cite{FukuiChern,FukuiZ2}

We have numerically benchmarked Eq.(\ref{eq: C1 and P2}) in the three-dimensional bulk pyrochlore iridates by evaluating the strong $Z_2$ topological index at $U=0,6,6.11$,\cite{Ara2012} which has eight time-reversal invariant momentum points (TRIM). The reciprocal lattice vector is written as
\begin{equation}
\label{eq: K vector}
\mathbf{K}=n_1 \mathbf{b}_1+n_2 \mathbf{b}_2+n_3 \mathbf{b}_3, \  \  n_{1,2,3} \in \mathbb{Z}
\end{equation}
Where $\mathbf{b}_1,\mathbf{b}_2,\mathbf{b}_3$ are basis vectors correspond to real lattice vectors $\mathbf{a}_1,\mathbf{a}_2,\mathbf{a}_3$. The strong topological index can be obtained by $P_3=(P_2(n_i=0)+P_2(n_i=1))\
\mathrm{mod}\ 2$ for $i=1,2,3$, where $P_2$ is evaluated from Eq.(\ref{eq:obstruction to Stokes}). With the results summarized in Table\ref{table:Z2}, one can verify the  strong topological insulator index $(1;000)$ in the 3D bulk pyrochlore iridates.

\begin{table}[h!]
\begin{tabular}{|c|c|c|}
\hline
& $P_2$ mod $2$ & $P_3$ mod $2$ \\
\hline
$n_1=0$&1&1 \\
\cline{1-2}
$n_1=1$&0& \\
\hline
$n_2=0$&1&1 \\
\cline{1-2}
$n_2=1$&0& \\
\hline
$n_3=0$&1&1 \\
\cline{1-2}
$n_3=1$&0& \\
\hline
\end{tabular}
\caption{The $Z_2$ index of the 3D bulk pyrochlore iridates at U$\le$6.11.}\label{table:Z2}
\end{table}


\section{Quasiparticle Effective Hamiltonian}
\label{sec:quasiham}

In this section, we derive the formalism for the quasiparticle effective Hamiltonian. A generic Green{'}s function can be written as
\begin{equation}
G_{\alpha \beta }(\omega, \mathbf{k})=\Big([(\omega +\mu +i \delta)\cdot \mathbf{1}-\mathbf{t}(\mathbf{k})-\mathbf{\Sigma }(\omega ,\mathbf{k})]^{-1}\Big){}_{\alpha
\beta}
\label{eq:G_generic}
\end{equation}
where $\alpha $, $\beta $ are some arbitrary quantum numbers (orbital, spin, or sites in a unit cell, etc), and $\mathbf{t}(\mathbf{k})$ is the Fourier transformed hopping integral. The self-energy is in general a complex matrix, but from the Lehman representation, we have
\begin{equation}
\label{eq:G_zero}
\mathbf{G}^{\dagger }(0,\mathbf{k})=\mathbf{G}(0,\mathbf{k}),
\end{equation}
and
\begin{equation}
\label{eq:Sigma_zero}
\mathbf{\Sigma }^{\dagger }(0,\mathbf{k})=\mathbf{\Sigma }(0,\mathbf{k}).
\end{equation}
At general complex frequency, we can separate the self-energy into Hermitian part and anti-Hermitian part as
\begin{eqnarray}
\mathbf{\Sigma }^H(z,\mathbf{k})&=&\left.\left(\mathbf{\Sigma }(z,\mathbf{k})+\mathbf{\Sigma }(z^*,\mathbf{k})^{\dagger }\right)\right/2 \\
\mathbf{\Sigma }^A(z,\mathbf{k})&=&\left.\left(\mathbf{\Sigma }(z,\mathbf{k})-\mathbf{\Sigma }(z^*,\mathbf{k})^{\dagger }\right)\right/2
\label{eq:Sigma_separation}.
\end{eqnarray}
The quasiparticle band structure can be defined as\cite{Okamoto2011}
\begin{equation}
\label{eq:quasi-dispersion}
\text{Det}\Big|(\omega +\mu )\cdot \mathbf{1}-\mathbf{t}(\mathbf{k})-\mathbf{\Sigma }^H(\omega ,\mathbf{k})\Big|\equiv 0.
\end{equation}
To analytically solve Eq.(\ref{eq:quasi-dispersion}) around the Fermi level, we expand the Hermitian part of the self-energy up to first order in $\omega $,
\begin{equation}
\label{eq:Sigma_expand}
\left.\mathbf{\Sigma }^H(\omega ,\mathbf{k})=\mathbf{\Sigma }^H(0,\mathbf{k})+\frac{\partial \mathbf{\Sigma }^H(\omega ,\mathbf{k})}{\partial \omega }\right
|_{\omega =0}\omega + \cdots
\end{equation}
Plugging the above into Eq.(\ref{eq:quasi-dispersion}), now we have
\begin{widetext}
\begin{equation}
\left. \text{Det} \Big| \omega \cdot \left(\mathbf{1}-\frac{\partial
\mathbf{\Sigma}^H(\omega ,\mathbf{k})}{\partial \omega } \right|_{\omega =0}\right) \\   -  \big(\mathbf{t}(\mathbf{k})-\mu \cdot \mathbf{1}+\mathbf{\Sigma}^{H}(\omega,\mathbf{k})\big) \Big|  \equiv 0.
\label{eq:quasi-dispersion-approx}
\end{equation}
\end{widetext}
Eq.(\ref{eq:quasi-dispersion-approx}) can  be converted into an eigenvalue problem of an artificial Hamiltonian. First we diagonalize the matrix to the first order of $\omega$ in Eq.(\ref{eq:quasi-dispersion-approx}) as
\begin{eqnarray}
B(\mathbf{k})&=&\left(
\begin{array}{ccc}
 \alpha _1(\mathbf{k}) & 0 & 0 \\
 0 & \alpha _2(\mathbf{k}) & 0 \\
 0 & 0 & \ddots \\
\end{array}
\right) \\ \nonumber
&=&U(\mathbf{k})\left(\mathbf{1}-\left.\left.\frac{\partial \mathbf{\Sigma }^H(\omega ,\mathbf{k})}{\partial \omega }\right.\right|_{\omega =0}\right)U^{\dagger
}(\mathbf{k}).
\label{eq:B(k)}
\end{eqnarray}
For simplicity, we drop the momentum index $\mathbf{k}$ in $B(\mathbf{k})$ and $U(\mathbf{k})$ [$U^{\dag}(\mathbf{k})$],
so that Eq.(\ref{eq:quasi-dispersion-approx}) can be rewritten as
\begin{eqnarray}
\text{Det}\Big| \omega \cdot B-U\left(\mathbf{t}(\mathbf{k})-\mu \cdot \mathbf{1}+\mathbf{\Sigma }^H(0,\mathbf{k})\right)U^{\dagger}
\Big|  \equiv 0. \nonumber
\label{eq:quasi-dispersion-approx2}
\end{eqnarray}
Let us rewrite
\begin{equation}
W=\left(
\begin{array}{ccc}
 \frac{1}{\sqrt{\alpha _1(\mathbf{k})}} & 0 & 0 \\
 0 & \frac{1}{\sqrt{\alpha _2(\mathbf{k})}} & 0 \\
 0 & 0 & \ddots \\
\end{array}
\right),
\label{eq:W(k)}
\end{equation}
then Eq.(\ref{eq:quasi-dispersion-approx}) becomes
\begin{eqnarray}
\text{Det}\Big| \omega \cdot \mathbf{1}-W
U\big(\pmb{t}(\mathbf{k})-\mu \cdot \mathbf{1}+\mathbf{\Sigma }^H(0,\mathbf{k})\big)U^{\dagger }W
\Big| \equiv 0. \nonumber
\label{eq:quasi-dispersion-approx3}
\end{eqnarray}
At this stage, we have defined the  ``quasiparticle dispersion" by solving the eigenvalue problem
\begin{equation}
H_{\text{eff}}(\mathbf{k})\psi ^{\alpha }(\mathbf{k})=E^{\alpha }(\mathbf{k})\psi ^{\alpha }(\mathbf{k}),
\label{eq:quasi-eigen}
\end{equation}
for the effective quasiparticle Hamiltonian,
\begin{equation}
H_{\text{eff}}(\mathbf{k})=WU\left(\mathbf{t}(\mathbf{k})-\mu \cdot \mathbf{1}+\mathbf{\Sigma }^H(0,\mathbf{k})\right)U^{\dagger}W.
\label{eq:quasi-ham}
\end{equation}
In the CDMFT formalism, since self-energy has no dependence on $k$,
\begin{equation}
\Sigma (0,\mathbf{k})\approx \Sigma (0),
\end{equation}
Eq.(\ref{eq:quasi-ham})  becomes
\begin{equation}
\label{eq:quasi-dispersion-CDMFT}
H_{\text{eff}}^{\text{CDMFT}}(\mathbf{k})=W U \left(\mathbf{t}(\mathbf{k})-\mu \cdot \mathbf{1}+\mathbf{\Sigma }^H(0)\right)U^{\dagger }W.
\end{equation}
The quasiparticle weight matrix is calculated by applying the Cauchy-Riemann equations
\begin{eqnarray}
& & \left(Z^{-1}\right)_{\alpha \beta } \\ \nonumber
&=& \left(\mathbf{1}-\left.
\frac{\partial \mathbf{\Sigma }^H(\omega
)}{\partial \omega }\right|_{\omega =0}\right)_{\alpha \beta } \\ \nonumber
& \approx &\delta _{\alpha \beta }-\frac{\text{Im}\left[\left(\mathbf{\Sigma }^A\left(\omega _0\right)\right)_{\alpha
\beta }\right]}{\omega _0}+i\frac{\text{Re}\left[\left(\mathbf{\Sigma }^A\left(\omega _0\right)\right)_{\alpha \beta }\right]}{\omega _0},
\end{eqnarray}
where $\omega_0=\pi/\beta$ is the first positive Matsubara frequency.
\end{document}